\documentclass[preprint,pra,onecolumn]{revtex4}
\usepackage{amsfonts}
\usepackage{amsmath}
\usepackage{amssymb}

\usepackage{fixme}
\usepackage{graphicx}
\usepackage{dcolumn}
\usepackage{bm}
\usepackage{color}

\begin{document}
\title{A general model on the simulation of the measurement-device independent quantum key distribution}
\author{Qin Wang$^{1}$}
\author{Xiang-Bin Wang$^{2,3}$}
\email{xbwang@mail.tsinghua.edu.cn}
\affiliation{$^{1}$Institute of Signal Processing and Transmission, Nanjing University of Posts and Telecommunications, Nanjing 210003, China}
\affiliation{$^{2}$Department of Physics and State Key Laboratory of Low Dimensional Quantum Physics, Tsinghua University, Beijing 100084, China}
\affiliation{$^{3}$Jinan Institute of Quantum Technology, Shandong Academy of Information Technology, Jinan, China}

\begin{abstract}

PACS number(s): 42.50.Ct, 78.67.Hc, 78.47.D-

We present a general model on the simulation of the measurement-device independent quantum key distribution (MDI-QKD). It can be used to predict experimental observations of a MDI-QKD with linear channel loss, simulating corresponding values for the gains, the error rates in different basis, and also the final key rates. Our model can be applicable to the MDI-QKDs with whatever convex source states or using whatever coding schemes. Therefore, it is useful in characterizing and evaluating the performance of any MDI-QKD protocols, making it a valuable tool in studying the quantum key distributions.

\end{abstract}

\maketitle

\section{Introduction}
There has been a long history between the attacks and the anti-attacks in the development of quantum key distributions (QKD) since the idea of BB84 (Bennett-Brassard 1984 \cite{BB84,GRTZ02}) protocol was put forward, due to the conflictions between the "in-principle" unconditional security and realistic implementations. Till today, there have been many different proposals for the secure QKD with realistic setups, such as the decoy-state method \cite{ILM,H03,wang05,LMC05,AYKI,qin1,qin2,haya,peng,wangyang,rep,njp} which can rescue the QKD with imperfect single-photon sources \cite{PNS1,PNS}, while the device-independent quantum key distribution \cite{ind1,gisin1} and the recently proposed measurement-device independent quantum key distribution (MDI-QKD) \cite{ind2,ind3} can further relieve the QKD even when the detectors are controlled by the eavesdropper \cite{lyderson}. Most interestingly, the MDI-QKD is not only immune to any detector attacks, but also able to generate a significant key rate with existing technologies. Moreover, its security can still be maintained with imperfect single-photon sources \cite{ind2,wangPRA2013,tittel1,liuyang,qin3,lopa,curtty}, and the effects of coding errors have also been studied \cite{wangPRA2013,kiyoshi}.

In developing practical QKDs, one important question is how to evaluate the performance of a proposal before really implementing it, since it is not realistic to experimentally test everything. Therefore, it is crucially important to make a thorough theoretical study and numerical simulation to predict the experimental results. In principle, it allows to use different kinds of sources in a decoy state MDI-QKD \cite{wangPRA2013,qin3}. Before experimentally testing all of them, one can choose to give a theoretical comparison with a reasonable model. In traditional decoy state methods \cite{H03,wang05,LMC05}, the models for calculation are relatively simple. However, for MDI-QKDs, it is not a simple job except for the special case of using weak coherent states. So far, there have been proposals with different sources, e.g., the heralded single-photon source (HSPS) \emph{etc} \cite{qin1,qin2,qin3}. And it has been shown that such a source can promise a longer secure distance than the weak coherent state. Nevertheless, it is unknown whether there are other sources which can present even better performance. Therefore, a general model on simulating the performance of arbitrary source states will be highly desirable. Here in this manuscript we solve the problem.

For simplicity, we assume a linear lossy channel in our model. Note that the security does not depend on the condition of linear loss at all. We only use this model to predict: what values the gains and error rates would possibly be observed if one did the experiment in the normal case when there is no eavesdropper. Given these values, one can then calculate the low bound of the yield and the upper bound of the phase flip-error rates for single-photon pairs. The major goal here is to simulate the values of gains and error rates of different states in normal situations. Of course, they can be replaced with the observed values in real implementations.

The paper is arranged as follows: In Sec. II we present the general model for the gains and error rates in a MDI-QKD, describing the detailed calculation processes. In Sec. III we proceed corresponding numerical simulations, comparing the different behaviors of MDI-QKDs when using different source states. Finally, discussions and summaries are given out in Sec. IV.

\section{The general model on MDI-QKD}

\begin{figure}[ptb]
\begin{center}
\includegraphics[scale=0.8]{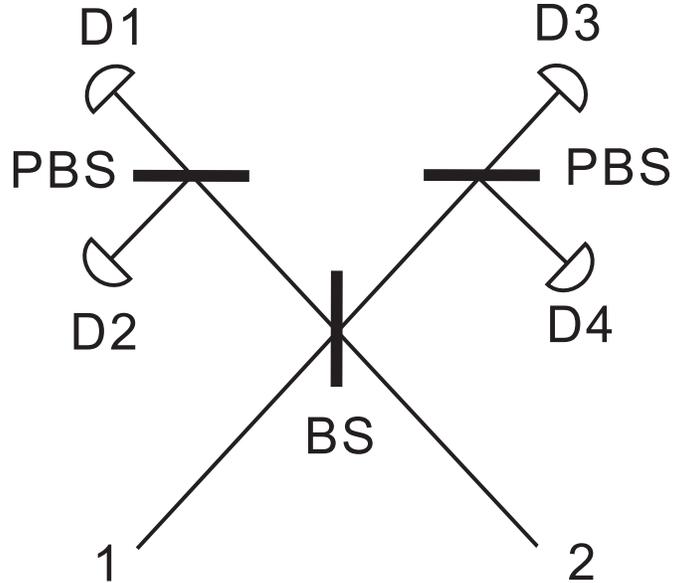}
\end{center}
\caption{(Color online) A schematic of the experimental setup for the collective measurements at the UTP. BS: beam-splitter; PBS: polarization beam-splitter; D1 - D4: single-photon detector; 1, 2: input port for photons. }
\label{Fig1}
\end{figure}

\subsection{Setups and definitions}
Consider the schematic setup in Fig. \ref{Fig1} \cite{ind2}, there are three parties, the users-Alice and Bob, and the un-trusted third party (UTP)-Charlie. Alice and Bob send their polarized photon pulses to the UTP who will take collective measurement on the pulse-pairs. The collective measurement results at the UTP determine the successful events. They are two-fold click of detectors (1,4), (2,3), (1,2) or (3,4). The gain of any (two-pulse) source is determined by the number of successful events from the source. There are 4 detectors at the UTP, we assume each of them has the same dark count rate $d$, and the same detection efficiency $\xi$. In such a case, we can simplify our model by attributing the limited detection efficiency to the channel loss. Say, if the actual channel transmittance from Alice to Charlie is $\eta_1$, we shall assume perfect detection efficiency for Charlie's detectors with channel transmittance of $\eta_1\xi$. Each detector will detect one of the 4 different modes, say $a_H^\dagger, a_V^\dagger, b_H^\dagger, b_V^\dagger $ in creation operator. For simplicity, we denote them by $c_i^\dagger$, i.e., $c_1^\dagger=a_H^\dagger, c_2^\dagger= a_V^\dagger, c_3^\dagger = b_H^\dagger, c_4^\dagger = b_V^\dagger$.
In such a way, detector $D_i$ corresponds to mode $i$ exactly. To calculate the gains that would-be observed for different source states in the linear lossy channel, we need to model the probabilities of different successful events conditional on different states. Let's first postulate some definitions before further study.

Definition 1: {\em event} $(i,j)$.
We define event $(i,j)$ as the event that both detector $i$ and detector $j$ click while other detectors do not click. Obviously, each $i,j$ must be from numbers $\{1,2,3,4\}$ and $i\not=j$ . For simplicity, we request $i<j$ throughout this paper.

Definition 2: {\em Output states and conditional probabilities of each events: notations} $\rho_{out}$: the output state of the beam-splitter. $|l_i,l_j\rangle = |l_il_j\rangle$: the beam-splitter's specific output state of $l_i$ photon in mode $i$, $l_j$ photon in mode $j$, and no photon in any other mode. Explicitly, $|l_il_j\rangle=\frac{1}{\sqrt {l_i!l_j!}}{c_i^\dagger}^{l_i} {c_j^\dagger}^{l_j}|0\rangle$. $P(ij|l_i,l_j)$ and $P(ij|\rho_{out})$:  the probability that event $(i,j)$ happens conditional on that the beam-splitter's output state is $|l_il_j\rangle$ and $\rho_{out}$, respectively. Hereafter, we omit the comma between $l_i$ and $l_j$, i.e., we use $|l_il_j\rangle$ for $|l_i,l_j\rangle$, and  $P(ij|l_il_j)$ for $P(ij|l_i,l_j)$.

Definition 3: {\em Events' probability conditional on the beam-splitter's input state:} $p_{ij}^{\alpha\beta}(k_1,k_2)=p_{ij}^{\alpha\beta}(k_1k_2)$. We denote
$p_{ij}^{\alpha\beta}(k_1,k_2)$ as the probability of event $(i,j)$ conditional on that there are $k_1$ photons of polarization $\alpha$ for mode $a$ and $k_2$ photons of polarization $\beta$ for mode $b$ as the input state of the beam-splitter. Hereafter, we omit the comma between $k_1$ and $k_2$. $\alpha$ or $\beta$ indicate the photon polarization. Explicitly, $\alpha$ or $\beta$ can be $H,V,+,-$ for polarizations of horizontal, vertical, $\pi/4$ and $3\pi/4$, respectively. To indicate the corresponding polarization state, we simply put each of these symbols inside a ket.

Definition 4: {\em Events' probability conditional on the two-pulse state of Alice and Bob's source:} $q_{ij}^{\alpha\beta}(\rho_A\otimes\rho_B)$.
It is the  probability that event $(i,j)$ happens conditional on that Alice sends out photon-number state $\rho_A$ with polarization $\alpha$ and Bob sends out photon number state $\rho_B$ with polarization $\beta$. Sometimes we simply use $q_{ij}^{\alpha,\beta}$ for simplicity.
\subsection{Elementary formulas and outline for the model}
Given the definitions above, we now formulate various conditional probabilities.
We start with the probability of event $(i,j)$ conditional on the output state $|l_il_j\rangle$.
\begin{eqnarray}\label{base}
P(ij|l_il_j)=\left\{
\begin{array}{l}
(1-d)^2,\; {\rm if}\; l_i>0, l_j>0\\
d(1-d)^2,\; {\rm if}\; l_i\cdot \l_j =0\; {\rm and}\; l_i+l_j >0\\
d^2(1-d)^2,\;  {\rm if}\; l_i=l_j=0
\end{array}\right.
\end{eqnarray}
Here the detection efficiency does not appear because we put shall this into the channel loss and hence we assume perfect detection efficiency. The factor $(1-d)^2$ comes from the fact that we request detectors other than $i,j$ {\em not} to click.  Also, the probability for event $(i,j)$ is 0 if any mode other than $i,j$ is not vacuum. Given these, we can now calculate probability distribution of the various two fold events given arbitrary input states of the beam-splitter. Therefore, for any output state of the beam-splitter $\rho_{out}$, the probability that event $(i,j)$ happens is
\begin{equation}\label{core}
P(ij|\rho_{out})= \sum_{l_i,l_j} P(ij|l_il_j)\langle l_il_i|\rho_{out}|l_il_j\rangle
\end{equation}
Based on this important formula, we can calculate the probability of event $(i,j)$ for any input state by this formula. For the purpose, we only need to formulate $\rho_{out}$. Therefore, given the source state of the two pulses $\rho_A\otimes\rho_B$,  we can use the following procedure to calculate the probability of event $(i,j)$, $p_{ij}(\rho_A\otimes\rho_B)$:
\\ i) Using the linear channel loss model to calculate the two-pulse state when arriving at the beam-splitter. Explicitly, if the channel transmittance is $\eta$, any state $|n\rangle\langle n|$ is changed into
\begin{equation}
\label{lin}|n\rangle\langle n|\longrightarrow \sum C_n^k \eta^k(1-\eta)^{n-k}|k\rangle\langle k|.
\end{equation}
\\ ii) Using the transformation: $ a_{H,V}^\dagger\longrightarrow \frac{1}{\sqrt 2} (a_{H,V}^\dagger + b_{H,V}^\dagger$; $b_{H,V}^\dagger =\frac{1}{\sqrt 2} (a_{H,V}^\dagger - b_{H,V}^\dagger)$ to calculate the output state of the beam-splitter, $\rho_{out}$.
\\ iii)  Using Eq.(\ref{core}) to calculate the probability of event $(i,j)$.
According to the protocol, we shall only be interested in the probabilities of successful events, $(1,2)$, $(3,4)$, $(1,4)$ and $(2,3)$. Below we will describe the detailed calculation processes in Z basis and X basis individually.

In Z basis, all successful events correspond to correct bit values when Alice and Bob send out {\em orthogonal} polarizations, and they correspond to wrong bit values when Alice and Bob send out the {\em same} polarizations. The observed gain in $Z$ basis for photon-number state $\rho_A\otimes\rho_B$ is,
\begin{equation}
S^Z_{\rho_A\otimes\rho_B} = \frac{1}{4}\sum_{(i,j)\in Suc} \left[
 q_{ij}^{HV}(\rho_A\otimes\rho_B) +
 q_{ij}^{VH}(\rho_A\otimes\rho_B) +
q_{ij}^{HH}(\rho_A\otimes\rho_B)+
q_{ij}^{VV}(\rho_A\otimes\rho_B) \right]
\end{equation}
and the set $Suc=\{(1,2), (3,4), (1,4), (2,3)\}$. Here, as defined in {Definition 4}, $q_{ij}^{\alpha\beta}(\rho_A\otimes\rho_B)$ represents the probability of event $(i,j)$ whenever Alice sends her photon number state $\rho_A$ with polarization $\alpha$ and Bob sends his photon number state $\rho_B$ with polarization $\beta$. For simplicity, we shall omit $\rho_A\otimes \rho_B$ in brackets or in subscripts if there is no confusion. Meantime, the successful events caused by the same polarizations will be counted as wrong bits. These will contribute to the bit-flip rate by:
\begin{equation}
\tilde E^Z = \frac{\sum_{(i,j)\in Suc} \left[
q_{ij}^{HH}+
q_{ij}^{VV}\right]}{4S_Z}
\end{equation}

In X basis, we should be careful that the situation is different from in Z basis, since now the successful events correspond to correct bits include two parts: 1) Alice and Bob send out the same polarizations ($++$ or $--$), and Charlie detects $\Phi^ + $ ((1,2) or (3,4) events happen); 2) Alice and Bob send out orthogonal polarizations ($+-$ or $-+$), and Charlie detects $\Psi ^ -$ ((1,4) or (2,3) events happen). And the left successful events belong to wrong bits. Therefore, we have
\begin{equation}
S^X = \frac{1}{4}\sum_{(i,j)\in Suc} \left[
 q_{ij}^{+-} +
 q_{ij}^{-+} +
q_{ij}^{++}+
q_{ij}^{--} \right]
\end{equation}
and
\begin{equation}
\tilde E^X = \frac{\sum_{(i,j)\in {(14),(23)}} \left[
q_{ij}^{++}+q_{ij}^{--}\right]+\sum_{(i,j)\in {(12),(34)}} \left[
q_{ij}^{+-}+q_{ij}^{-+}\right]}{4S_X}
\end{equation}

Moreover, there are alignment errors which will cause a fraction ($E_{d}$) of states to be flipped. We then modify the error rate in different bases by
\begin{equation}\label{alin1}
E^Z = E_{d}\cdot (1-2 \tilde E^Z) + \tilde E^Z
\end{equation}
and
\begin{equation}\label{alin2}
E^X = E_{d}\cdot (1-2\tilde E^X) +\tilde E^{X}
\end{equation}

Note that in the above two formulas above, we have considered this fact: before taking the alignment error into consideration, the successful events can be classified into two classes: one class has no error and the other class has an error rate of $50\%$, they are totally random bits. The second class takes a fraction of $2E^Z$ (or $2E^X$) among all successful events. Alignment error does not change the error rate of the second class of events, since they are random bits only.

Given these, we can simulate the final key rate. In a model of numerical simulation, our goal is to deduce the probably would-be value for $S^Z, S^X$ and $E^Z,E^X$ in experiments. Given these, one can then calculate the yield of the single-photon pairs, $s_{11}$, the bit-flip rates in $Z$ basis and $X$ basis, and hence the final key rate. Now everything is reduced to calculate all $p_{ij}^{\alpha\beta}$ above.

\subsection{Conditional probabilities for beam-splitter's incident state of $k_1$ photons in mode $a$ and $k_2$ photons in mode $b$}
We consider the case that there are $k_1$ incident photons in mode $a$ and $k_2$ incident photons in mode $b$ of the beam-splitter. Each incident pulse of the beam-splitter has its own polarization and is indicated by a subscript. In general, we consider the state
\begin{equation}
|k_1\rangle_\alpha|k_2\rangle_\beta
\end{equation}
 We shall consider the conditional probabilities for various successful events, i.e. $p_{ij}^{\alpha\beta}(k_1k_2)$. Since we only consider the incident state of $k_1$ photons in mode $a$ and $k_2$ photons in mode $b$, we shall simply use $p_{ij}^{\alpha\beta}$ for $p_{ij}^{\alpha\beta}(k_1k_2)$ in what follows.

i) in Z basis

First, we consider the following two-mode state
\begin{equation}
|k_1\rangle_H|k_2\rangle_V
=\frac{1}{\sqrt{k_1!k_2!}}{a_H^\dagger}^{k_1}{b_V^\dagger}^{k_2}|0\rangle
\end{equation}
as the input state of the beam-splitter.
After BS, the output state $|\psi\rangle$ is
\begin{equation}
|\psi\rangle=\left(\frac{1}{\sqrt{2}}\right)^{k_1+k_2}\frac{1}{\sqrt{k_1!k_2!}}(a^\dagger_H+b^\dagger_H)^{k_1}
(a^\dagger_V-b^\dagger_V)^{k_2}|0\rangle
\end{equation}
Therefore
\begin{equation}
\langle l_1l_2|\rho_{out}|l_1l_2 \rangle =(1/2)^{k_1+k_2}\delta_{k_1l_1}\delta_{k_2l_2}
\end{equation}
According to Eq.(\ref{core}), the conditional probability for event (1,2) is
\begin{equation}
p_{12}^{HV} = P(12|\rho_{out})
=\sum_{l_1,l_2} P(12|l_1l_2)(1/2)^{k_1+k_2}\delta_{k_1l_1}\delta_{k_2l_2}
=(1/2)^{k_1+k_2} P(12|k_1k_2)
\end{equation}
Similarly, we have
\begin{eqnarray}
\begin{array}{l}
p_{34}^{HV} =(1/2)^{k_1+k_2} P(34|k_1k_2)\\
p_{14}^{HV} =(1/2)^{k_1+k_2} P(14|k_1k_2)\\
p_{23}^{HV} =(1/2)^{k_1+k_2} P(23|k_2k_1)
\end{array}
\end{eqnarray}
Note that here $P(ij|k_mk_n)$ is just $P(ij|l_i=k_m,l_j=k_n)$ when $l_1=k_1$ as defined by our Definition 2 in previous section. For example, $P(23|k_2k_1)$ is actually $P(23|l_2=k_2, l_3=k_1)$. Similarly, if the beam-splitter's input state is $|k_1\rangle_V|k_2\rangle_H$, i.e. $k_1$ vertical photons in mode $a$ and $k_2$ horizontal photons in mode $b$, we have
\begin{eqnarray}\begin{array}{l}
p_{12}^{VH}=(1/2)^{k_1+k_2} P(12|k_2k_1)\\
p_{34}^{VH}=(1/2)^{k_1+k_2} P(34|k_2k_1)\\
p_{14}^{VH}=(1/2)^{k_1+k_2} P(14|k_2k_1)\\
p_{23}^{VH}=(1/2)^{k_1+k_2} P(23|k_1k_2)
\end{array}\end{eqnarray}
Next we consider the following two-mode state
\begin{equation}
|k_1\rangle_H|k_2\rangle_H
=\frac{1}{\sqrt{k_1!k_2!}}{a_H^\dagger}^{k_1}{b_H^\dagger}^{k_2}|0\rangle
\end{equation}
as the input state of the beam-splitter. After the beam-splitter, it changes into:
\begin{equation}
|\psi\rangle =
\left(\frac{1}{\sqrt{2}}\right)^{k_1+k_2}\frac{1}{\sqrt{k_1!k_2!}}(a^\dagger_H+b^\dagger_H)^{k_1}
(a^\dagger_H-b^\dagger_H)^{k_2}|0\rangle
\end{equation}
We have the following uniform formula for probabilities of any successful events:
\begin{eqnarray}
p_{ij}^{HH} =\left\{\begin{array}{l}\frac{(k_1+k_2)!}{k_1!k_2!}(1/2)^{k_1+k_2}P(ij|k_1+k_2,0); \;
{\rm for}\; i=1,\;j=2; {\rm or}\; i=3,\;j=4\\
\frac{(k_1+k_2)!}{k_1!k_2!}(1/2)^{k_1+k_2}P(ij|0,k_1+k_2); \;
{\rm for}\; i=1,\;j=4; {\rm or}\; i=2,\;j=3\\
\end{array}\right.
\end{eqnarray}

Similarly, when the beam-splitter's input pulses are both vertical, we can find the value for $p_{ij}^{VV}$.

ii) in $X$ basis

We first consider the beam-splitter's input state of $|k_1\rangle_+|k_2\rangle_-$, i.e., there are $k_1$ photon with $\pi/4$ polarization in mode $a$ and $k_2$ photons with $3\pi/4$ polarization in mode $b$. Note that $|\pm\rangle=\frac{1}{\sqrt 2}(|H\rangle\pm|V\rangle)$. The output state of the beam-splitter is
\begin{equation}
|\psi\rangle = \frac{1}{2^{k_1+k_2}\sqrt{k_1!k_2!}} (a_H^\dagger + a_V^\dagger + b_H^\dagger  + b_V^\dagger)^{k_1} (a_H^\dagger - a_V^\dagger-b_H^\dagger + b_V^\dagger)^{k_2}|0\rangle
\end{equation}
We have
\begin{equation}
\langle l_il_j| \psi\rangle
=\frac{1}{2^{k_1+k_2}\sqrt{k_1!k_2!}}
\sum_{s=\Delta_1}^{\Delta_2}
\sqrt{l_i!l_j!} C_{k_1}^{s}C_{k_2}^{l_1-s}(-1)^{k_2-l_i+s}\delta_{l_i+l_2,k_1+k_2}
\end{equation}
where \begin{equation}\Delta_1=min\{l_i,k_1\},\;\; \Delta_2=l_i-min\{l_i,k_2\}\end{equation}and $min\{l_i,k_1(k_2)\}$ is the smaller one of $l_i$ and $k_1$($k_2$).
Thus we can calculate the conditional probabilities by
$$
p_{ij}^{+-} = \sum_{l_i=0}^{k_1+k_2} |\langle l_il_j| \psi\rangle|^2
$$
Hence
\begin{equation}
p_{ij}^{+-}=
\frac{1}{4^{k_1+k_2}k_1!k_2!}\sum_{l_i=0}^{k_1+k_2}\left|
\sum_{s=\Delta_2}^{\Delta_1}
\sqrt{l_1!(k_1+k_2-l_i)!} C_{k_1}^{s}C_{k_2}^{l_i-s}(-1)^{l_i-s}\right|^2P(ij|l_i,k_1+k_2-l_i)
\end{equation}
for $i=1,j=2$ and $i=3,j=4$;
and
\begin{equation}
p_{ij}^{+-}=\frac{1}{4^{k_1+k_2}k_1!k_2!}\sum_{l_i=0}^{k_1+k_2}\left|
\sum_{s=\Delta_2}^{\Delta_1}
\sqrt{l_i!(k_1+k_2-l_1)!} C_{k_1}^{s}C_{k_2}^{l_i-s}\right|^2P(ij|l_i,k_1+k_2-l_i)
\end{equation}
for $i=1,j=4$ and $i=2,j=3$. Besides, it is easy to show
\begin{equation}
p_{ij}^{-+} = p_{ij}^{+-}
\end{equation}
If the polarization of incident pulses of the beam-splitter are both $\pi/4$, then the output state is
\begin{equation}
|\psi\rangle = \frac{1}{2^{k_1+k_2}\sqrt{k_1!k_2!}} (a_H^\dagger + a_V^\dagger + b_H^\dagger  + b_V^\dagger)^{k_1} (a_H^\dagger +a_V^\dagger-b_H^\dagger - b_V^\dagger)^{k_2}|0\rangle.
\end{equation}
We find
\begin{equation}
p_{ij}^{++}
=\frac{1}{4^{k_1+k_2}k_1!k_2!}\sum_{l_i=0}^{k_1+k_2}\left|
\sum_{s=\Delta_2}^{\Delta_1}
\sqrt{l_1!(k_1+k_2-l_i)!} C_{k_1}^{s}C_{k_2}^{l_i-s}\right|^2P(ij|l_i,k_1+k_2-l_i)
\end{equation}
for $i=1,j=2$ and $i=3,j=4$;
and
\begin{equation}
p_{ij}^{++}=\frac{1}{4^{k_1+k_2}k_1!k_2!}\sum_{l_i=0}^{k_1+k_2}\left|
\sum_{s=\Delta_2}^{\Delta_1}
\sqrt{l_i!(k_1+k_2-l_1)!} C_{k_1}^{s}C_{k_2}^{l_i-s}(-1)^{l_i-s}\right|^2P(ij|l_i,k_1+k_2-l_i)
\end{equation}
for $i=1,j=4$ and $i=2,j=3$.
Also, we have
\begin{equation}
p_{ij}^{--}=p_{ij}^{++}
\end{equation}
\subsection{Probabilities of events conditional on source states }
In the above subsection, we have formulated the probabilities of various events conditional on a pure input state $|k_1\rangle|k_2\rangle$. In fact, the results can be easily extended to the more general case when the beam-splitter's input state is a mixed state. Say,
\begin{equation}
\label{bss}\left(\sum_{k_1}f_{k_1}|k_1\rangle\langle k_1\right)\otimes \left(\sum_{k_2}f_{k_2}|k_2\rangle\langle k_2|\right)
\end{equation}
Suppose the polarizations of mode $a,b$ are $\alpha,\beta$, respectively. We then have
\begin{equation}\label{inp}
p_{ij}^{\alpha\beta} = \sum_{k_1,k_2} f_{k_1}f_{k_2} p_{ij}^{\alpha\beta}(k_1k_2)
\end{equation}
where $p_{ij}^{\alpha\beta}(k_1k_2)$ is the same as defined in the previous subsection, for all possible polarizations $(\alpha,\beta) =(H,V),(V,H),(H,H),(V,V),(+,-),(-,+),(+,+),(-,-)$. To formulate the probabilities conditional on any source states, we only need to relate the source state with the beam-splitter's input state. Suppose the source state in photon-number space is $\rho_A\otimes\rho_B$ and
\begin{eqnarray}\label{sou}\begin{array}{l}
\rho_A = \sum_n a_n |n\rangle\langle n|\\
\rho_B = \sum_n b_n |n\rangle\langle n|
\end{array}
\end{eqnarray}
After some loss channel, the state changes into the beam-splitter's input state as Eq.(\ref{bss}). Suppose the transmittance for the channel between Alice (Bob) and UTP is $\eta_A$ ($\eta_B$).
Using the linear loss model of Eq. (\ref{lin}) we have
\begin{eqnarray}\label{ff}\begin{array}{l}
f_{k_1} = \sum_{n\ge k_1} a_n \eta_A^{k_1}(1-\eta_A)^{n-k_1}C_n^{k_1}\\
f_{k_2} = \sum_{n\ge k_2} b_n \eta_B^{k_2}(1-\eta_B)^{n-k_2}C_n^{k_2}
\end{array}
\end{eqnarray}
We now arrive at our major conclusion:

{\em Major conclusion:} Formulas of $p_{ij}^{\alpha\beta}(k_1k_2)$ in the earlier subsection together with Eqs. (\ref{inp},\ref{ff}) complete the model of probabilities of different events conditional on any source states, i.e., the gains. Using Eqs. (\ref{alin1},\ref{alin2}), one can also model the observed error rates of any source states.

\subsection{3-intensity decoy-state MDI-QKD}
Using the {\em Major conclusion} above, we can model the gains and the error rates with a 3-intensity decoy-state MDI-QKD method \cite{wangPRA2013,qin3}.
We assume that Alice (Bob) has three intensities in their source states, denoted as $0,\mu_A,\mu_A'$ ($0,\mu_B,\mu_B'$). Denote $\rho_x$ ($\rho_y$) as the density operator for source $x$ ($y$) at Alice's (Bob's) side,  and $x$ ($y$) can take any value from $0,\mu_A,\mu_A'$ ($0,\mu_B,\mu_B'$).
\begin{equation}\label{state0}
\rho_0 = |0\rangle\langle 0|;\\
\rho_{\mu_A}=\sum_{k} a_k |k\rangle\langle k|;\,\,\rho_{\mu_A'}=\sum_{k} a_k' |k\rangle\langle k|;\\
\rho_{\mu_B}=\sum_{k} b_k |k\rangle\langle k|;\,\,\rho_{\mu_B'}=\sum_{k} b_k' |k\rangle\langle k|
\end{equation}
Then we have the expression for the low bound of the yield of single-photon pulse pairs
\begin{equation}\label{y111}
Y_{11}^X \ge {Y_{11}^{X,L}}  \equiv  \frac{a_1'b_2'(S_{\mu,\mu}^{X} -\tilde S_0^{X} ) - a_1b_2(S_{\mu',\mu'}^{X} - \tilde S_0'^{X} )}{a_1'a_1(b_2'b_1-b_2b_1')}
\end{equation}
and their upper bound of the phase flip-error rate
\begin{equation}
e_{11}^{X} \leqslant e_{11}^{X,U}  \equiv  \frac{{E_{\mu ,\mu }^{X} S_{\mu ,\mu }^{X}  - E_{\mu ,0}^{X} S_{\mu ,0}^{X}  - E_{0,\mu}^{X} S_{0,\mu}^{X}+ E_{0,0}^{X} S_{0,0}^{X} }}
{{Y_{11}^{X} }}
\end{equation}
With the results above, now we can calculate the key rate with the formula \cite{ind2,wangPRA2013,qin3}
\begin{equation}
R \ge a_1^\prime b_1^\prime Y_{11}^{Z} [1 - H(e_{11}^{X} )] - S_{\mu '\mu '}^Z f(E_{\mu '\mu '}^Z )H(E_{\mu '\mu '}^Z )
\end{equation}

\section{Numerical simulations}
Using all the above correspondence, we can numerically simulate the gains and error rates of any source states. Taking as an example, we consider the source of a HSPS from parametric down-conversion processes \cite{qin3}. It originally has a Poissonian photon number distribution
when pumped by a continuous wave (CW) laser \cite{explain1}, written as:
\begin{equation}
\left| \psi  \right\rangle {\text{ = }}\frac{{x^n }}{{n!}}e^{ - x}
\end{equation}
where $x$ is the the average intensity of the emission light.
However, after chosen a proper gating time and triggered with a practical single photon detector, a sub-Poissonian distributed source state can be obtained, which can be expressed as:
\begin{eqnarray}\label{rho}
\begin{array}{l}
\rho  = [{\text{P}}^{{\text{Cor}}} d_i  + (1 - {\text{P}}^{{\text{Cor}}} )e^{ - x} ]\left| 0 \right\rangle \left\langle 0 \right| + \sum\limits_{n = 1}^\infty  {{\text{[P}}^{{\text{Cor}}} e^{ - x} \frac{{x^{n - 1} }}
{{(n - 1)!}} + (1 - {\text{P}}^{{\text{Cor}}} )e^{ - x} \frac{{x^n }}
{{n!}}]} \left| n \right\rangle \left\langle n \right|
\end{array}
\end{eqnarray}
where ${\text{P}}^{{\text{Cor}}}$ is the correlation rate of photon pairs, i.e., the probability that we can predict the existence of a heralded photon when a heralding one was detected;
$d_i$ is the dark count rate of the triggering detector.

\begin{figure}[ptb]
\begin{center}
\includegraphics[scale=0.8]{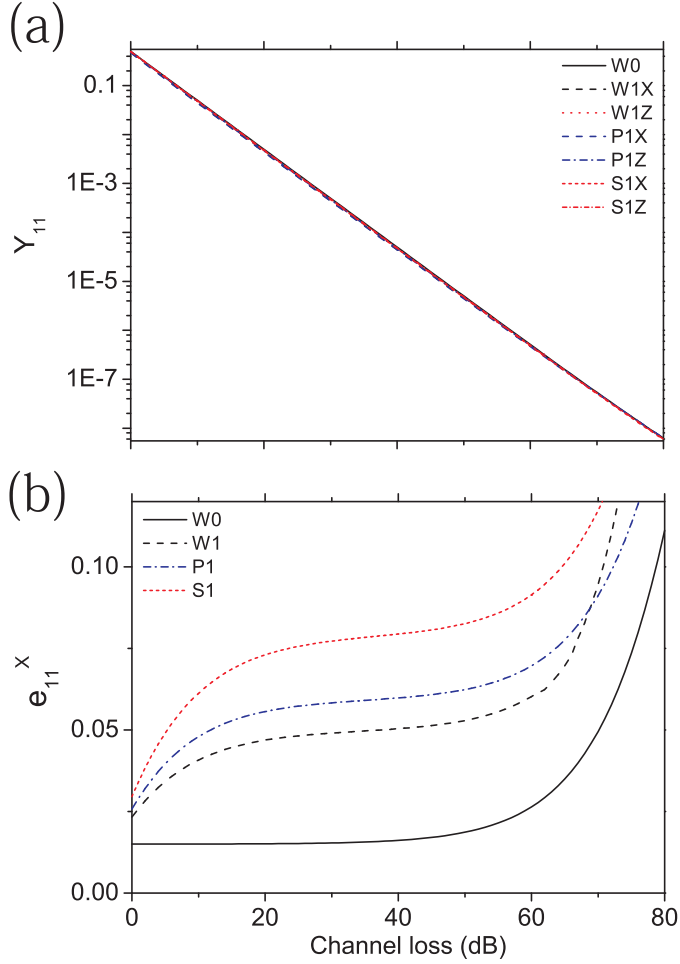}
\end{center}
\caption{(Color online) (a) The lower bound of $Y_{11}$ and (b) the upper bound of $e_{11}^X$ for different photon sources. The solid lines (W0) represent the results of using infinite-decoy state method, and the dashed or dotted lines (W1, P1 or S1) represent using three-decoy state method. Besides, W, P or S each corresponds to the scheme of using weak coherent sources \cite{ind2}, possonian heralded single photon sources \cite{qin2} or sub-possonian heralded single photon sources \cite{qin3}, individually. X or Z represent in X or Z basis respectively. Here at each point, we set $\mu = 0.05$, and optimize the value for $\mu'$.}
\label{Fig2}
\end{figure}
\begin{figure}[ptb]
\begin{center}
\includegraphics[scale=0.8]{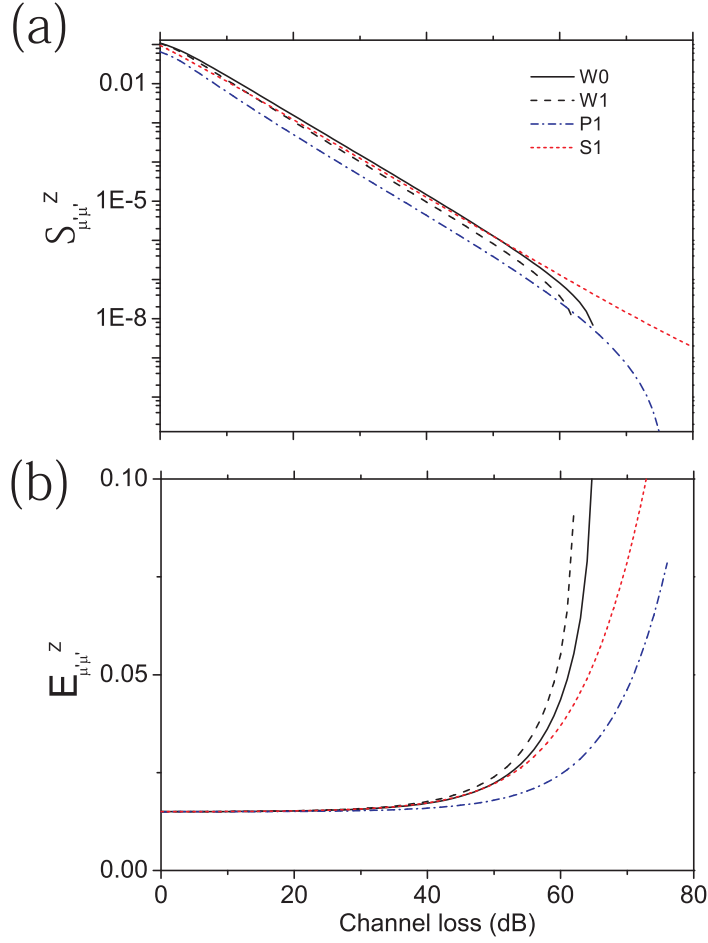}
\end{center}
\caption{(Color online) (a) The gain and (b) the quantum error-bit rate in Z basis for different photon sources. The solid lines (W0) represent the results of using infinite-decoy state method, and the dashed or dotted lines (W1, P1 or S1) represent using three-decoy state method. Besides, W, P or S each corresponds to the scheme of using weak coherent sources, possonian heralded single photon sources \cite{qin2} or sub-possonian heralded single photon sources \cite{qin3}, individually. Here at each point, we set $\mu = 0.05$, and optimize the value for $\mu'$.}
\label{Fig3}
\end{figure}
\begin{figure}[ptb]
\begin{center}
\includegraphics[scale=0.7]{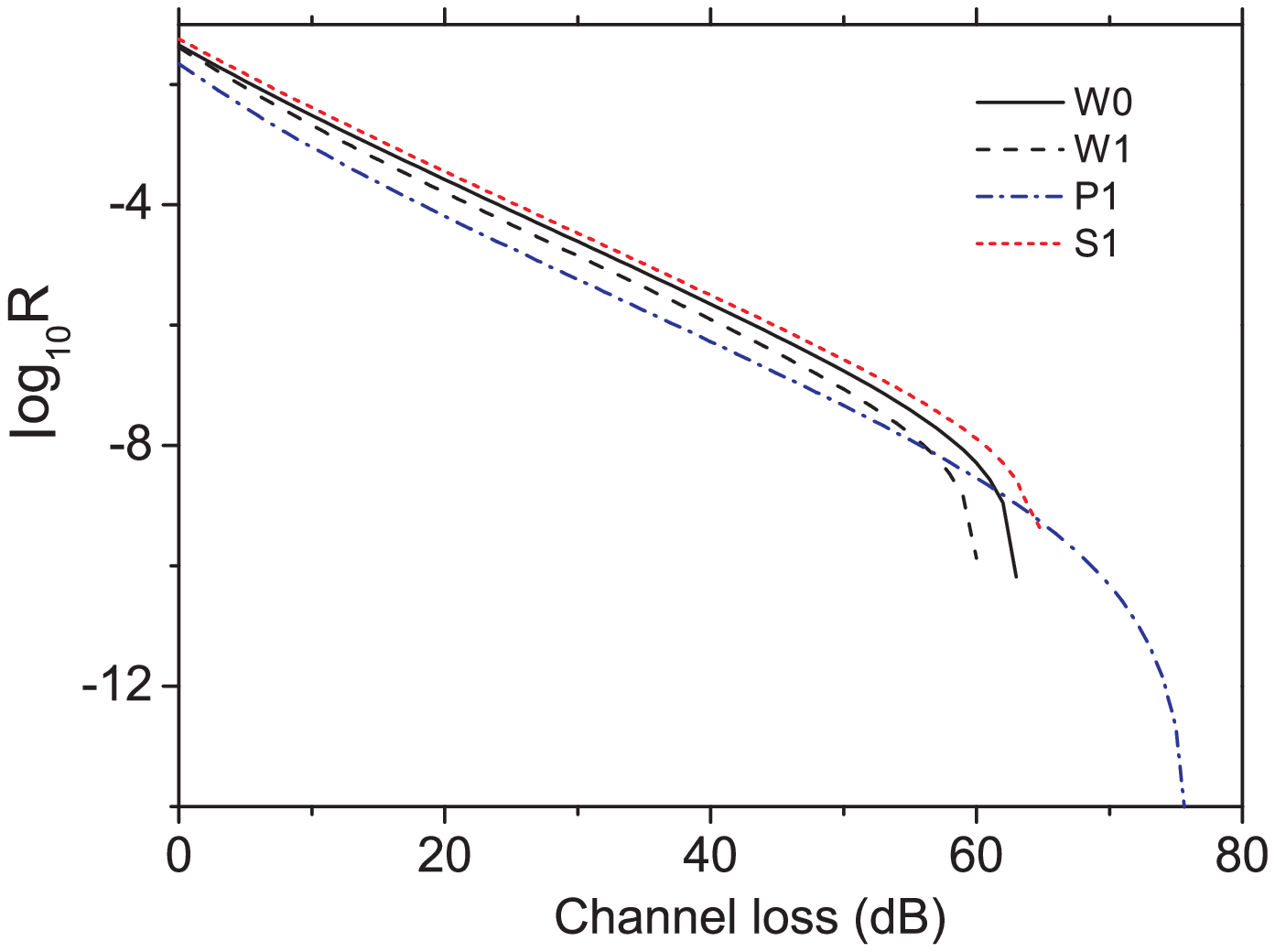}
\end{center}
\caption{(Color online) (a) The final key rate for different photon sources. The solid lines (W0) represent the results of using infinite-decoy state method, and the dashed or dotted lines (W1, P1 or S1) represent using three-decoy state method. Besides, W, P or S each corresponds to the scheme of using weak coherent sources, possonian heralded single photon sources \cite{qin2} or sub-possonian heralded single photon sources \cite{qin3}, individually. Here at each point, we set $\mu = 0.05$, and optimize the value for $\mu'$.}
\label{Fig4}
\end{figure}

In the following numerical simulations, for simplicity, we assume the UTP lies in the middle of Alice and Bob, and all triggering detectors (at Alice or Bob's side) have the same detection efficiency ($75\%$) and the same dark count rate ($10^{-6}$). We also assume all triggered detectors (at the UTP's side) have the same detection efficiency (they are attributed into the channel loss), and the same dark count rate ($3\times 10^{-6}$). Besides, we set the system misalignment probability to be $1.5\%$.

Fig. 2(a) and (b) each show the low bound of $Y_{11}$ (in X or Z basis) and the upper bound of $e_{11}^X$ changing with channel loss for different source states, i.e., the weak coherent sources (W), the possonian heralded single photon sources (P) and the sub-possonian heralded single photon sources (S). The solid line represents the result of using infinite number of decoy state method (W0), and the dashed or dotted lines (P1 or S1) are the results of using three-decoy state method.

Similar to Fig. 2(a) and (b), Fig. 3(a) and (b) each show corresponding values of the gains (${\text{S}}_{\mu '\mu '}^Z$) and the quantum bit-error rates (QBER) (${\text{E}}_{\mu '\mu '}^Z$) of signal pulses in Z basis for different source states. And Fig. 5 presents the final key rate changing with channel loss.

See from Fig. 4, we find that the sub-possonian heralded single photon sources can generate the highest key rate at lower or moderate channel loss ($\leqslant 64$ dB). Because within this range, its signal state has a lower QBER than in the weak coherent sources, and a higher gain than in the possonian heralded single photon sources as simulated in Fig. 3 (a) and (b). However, at larger channel loss ($\geqslant 64$ dB), the possonian heralded single photon source shows better performance than the other two, this is mainly due to its much lower vacuum component which may play an essential role in the key distillation process when suffering from lager channel loss.

\section{Conclusions}
In summary, we have presented a general model for simulating the gains, the error rates and the key rates for MDI-QKDs, which can be applicable to the schemes of using arbitrary convex source states and any coding methods. This facilitates the performance evaluation of any MDI-QKD methods, and thus make it a valuable tool for devising high efficient QKD protocols and for studying long distance quantum communications.

\section{ACKNOWLEDGMENTS}
We gratefully acknowledge the financial support from the National High-Tech Program of China through Grants No. 2011AA010800 and No. 2011AA010803, the NSFC through Grants No. 11274178, No. 11174177, No. 60725416 and No. 11311140250, and the 10000-Plan of Shandong province. The author-X. B. Wang thanks Y. H. Zhou and Z. W. Yu for useful discussion.


\begin{thebibliography}{99}
\bibitem{explain1} During a spontaneous parametric down conversion (SPDC) process, if a pulsed pump laser is used, as long as the coherence time of the emission, $\Delta t_c$, is much longer than the duration of the pump pulse, $\Delta t$, i.e., $\Delta t_c  \gg \Delta t$, (in practice easily obtained by using ultrafast (fs) pulse pump lasers), a single emission process will take place, giving an thermal photon number distribution. In contrast, when a continuous wave (CW) laser is used, as long as $\Delta t_c$ is much shorter than the gating period of the detector, a large number of independent SPDC processes will be present, each thermally distributed, but collectively resulting in a Poisson distribution. However, the "original" distribution can be altered by conditional gating. By choosing proper gating time and using an appropriate correlation rate, a sub-Poissonian distributed HSPS can be obtained as the result of postselections.
\end{thebibliography}
\end{document}